\documentclass{revtex4}

\usepackage{graphicx}

\begin{document}

\title{Dynamic Regimes in Films with a Periodic Array of Antidots}

\author{A. V. Silhanek}
\address{Nanoscale Superconductivity and Magnetism Group, Laboratory for Solid State Physics and Magnetism
K. U. Leuven\\ Celestijnenlaan 200 D, B-3001 Leuven, Belgium}

\author{S. Raedts}
\address{Nanoscale Superconductivity and Magnetism Group, Laboratory for Solid State Physics and Magnetism
K. U. Leuven\\ Celestijnenlaan 200 D, B-3001 Leuven, Belgium}

\author{M. J. Van Bael}
\address{Nanoscale Superconductivity and Magnetism Group, Laboratory for Solid State Physics and Magnetism
K. U. Leuven\\ Celestijnenlaan 200 D, B-3001 Leuven, Belgium}

\author{V.~V.~Moshchalkov}
\address{Nanoscale Superconductivity and Magnetism Group, Laboratory for Solid State Physics and Magnetism
K. U. Leuven\\ Celestijnenlaan 200 D, B-3001 Leuven, Belgium}

\date{\today}

\begin{abstract}
We have studied the dynamic response of Pb thin films with a square array of antidots by means of ac susceptibility $\chi(T,H)$ measurements. At low enough ac drive amplitudes $h$, vortices moving inside the pinning potential give rise to a  frequency- and $h$-independent response together with a scarce dissipation. For higher amplitudes, the average distance travelled by vortices surpasses the pinning range and a critical state develops. We found that the boundary $h^*(H,T)$ between these regimes smoothly decreases as $T$ increases whereas a step-like behavior is observed as a function of field. We demonstrate that these steps in $h^*(H)$ arise from sharp changes in the pinning strength corresponding to different vortex configurations. For a wide set of data at several fields and temperatures in the critical state regime, we show that the scaling laws based on the simple Bean model are satisfied. 
\end{abstract}


\maketitle

\section{Introduction}

AC susceptibility $\chi=\chi^{\prime} + i \chi^{\prime \prime}$ measurements are a powerful tool widely used for investigating the vortex dynamics in the mixed state of Type-II superconductors. From the experimental point of view, this inductive technique has the advantage of being inexpensive, highly sensitive, and very simple. On top of that, it allows one to access and probe different vortex regimes and phases by changing either the ``sensing parameters", like frequency $f$, amplitude $h$ and wave shape of the alternating excitation, or the thermodynamic variables like the external field $H$ and temperature $T$. For example, in the vortex solid phase, small values of the external excitation $h$ induce an oscillatory motion of the flux lines inside the pinning potential characterized by a very small dissipation and a response $\chi$ independent of $h$.\cite{campbell} In this so-called Campbell regime, it is possible to determine the average curvature of the pinning potential $\alpha$ assuming a parabolic potential well.\cite{silhanek03,review} 

At higher amplitudes $h$, the vortex displacement approaches the pinning range and the pinning potential departs from the harmonic approximation. At this stage, a nonlinear response is observed and the screening $\chi^\prime$ becomes $h$- and $f$-dependent. Eventually, at high enough amplitudes, the average distance travelled by vortices may surpass the distance between nearby pinning centers (inter-valley motion) and a critical state (CS) develops.\cite{review}

Unlike the Campbell regime, where the vortex configuration is preserved after measuring, in the critical state regime the initial vortex configuration may be strongly altered during the measurement process. However, it is precisely this invasive character of the measurements in the CS which allows one to determine the critical current density $J$ (i.e. the depth of the potential well), thus providing complementary information to that obtained in the Campbell regime (i.e. the curvature of the pinning potential). 

Typically the transition between the Campbell and the critical state regimes is broadened by topological and energetic disorder. These effects can be substantially reduced by introducing a regular array of pinning centers. In a recent work\cite{silhanek03} we have shown that this reduction of disorder leads, within the Campbell regime, to well defined transitions in the ac response at the commensurability fields $H=H_n=n \Phi_0 / d^2$, where $\Phi_0$ is the superconducting flux quantum and $d$ is the period of the square pinning lattice.

It is generally thought that the vortex distribution in this kind of samples consists of terraces of homogeneous vortex distribution connected by narrow walls where the vortex density changes abruptly.\cite{cooley} Within this picture, the applicability of standard critical state models with smooth field profiles is highly questionable, and even more doubtful is the validity of the simple Bean critical state model commonly used to extract the critical current from magnetization measurements. However, the lack of any corroborative evidence of such terraced critical state profile in samples with periodic pinning stimulates further experimental studies to gain insight into this rather unexplored topic. 

In this work we study the ac dynamic response of Pb thin films with a square pinning array of holes. We determine the crossover field from linear to nonlinear regimes as a function of temperature and field. We found that this boundary smoothly decreases with increasing temperature and exhibits a step-like field dependence with clear jumps at every matching field. For large amplitudes, in the critical state regime, the scaling relations derived from a simple Bean model are accurately satisfied for a large range of fields and temperatures. This analysis allows us to estimate quantitatively the temperature and field dependence of the critical current density as well as to determine the onset of the critical state regime. 

\section{Experimental Details}

The experiments were conducted on two Pb thin films with a square antidot array of period $d=1.5 \mu$m, which corresponds to a first matching field $H_1=9.2$ G. The antidots have a square shape with a size $b=0.8 \mu$m. The name, critical temperature and thickness $\delta$ of the used samples are, AD15 ($\delta=13.5$ nm, $T_c=7.1$ K) and AD65 ($\delta=65$ nm, $T_c=7.21$ K), respectively. From the $T_c(H)$ slope we have estimated, for both samples, a superconducting coherence length $\xi(0) \sim 33 \pm 1$ nm. 

The predefined resist-dot patterns were prepared by electron-beam lithography in a polymethyl metacrylate/methyl metacrylate (PMMA/MMA) resist bilayer covering the SiO$_2$ substrate. The use of two different resist layers is necessary to obtain an overhang profile which eventually guarantee that there will be no contact between the material deposited on the substrate and the resist layer. A Ge(20~\AA)/Pb/Ge(200~\AA) film was then electron-beam evaporated onto this mask while keeping the substrate at liquid nitrogen temperature. Finally, the resist was removed in a liftoff procedure in warm acetone. 

The ac-measurements were carried out in a commercial Quantum Design-PPMS device with drive field amplitudes $h$ ranging from 2 mOe to 10 Oe, and the frequency $f$ from 10 Hz to 15 kHz. In this window of frequencies we have found that $\chi$ is weakly dependent on $f$ and therefore we report results obtained at a single value $f=3837$ Hz. In all cases, the data were normalized by the same factor corresponding to a total step $\Delta \chi^{\prime} =$ 1 with $H=$0.

\section{Boundary of the linear regime}

\begin{figure}[htb]
\centering
\includegraphics[angle=0,width=90mm]{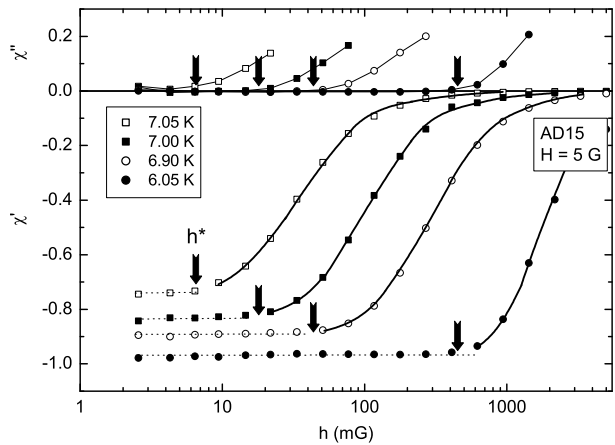}
\caption[]{{\small Ac-susceptibility $\chi=\chi^{\prime} + i \chi^{\prime \prime}$ as a function of the drive field $h$ for several temperatures at $H=5$ G. Horizontal dashed lines indicate the range of the linear response and the arrows the onset of the nonlinear regime.}}
\label{Xvshac}
\end{figure}

In order to identify the different dynamic regimes, we carried out a series of measurements of $\chi=\chi^{\prime} + i \chi^{\prime \prime}$ as a function of the ac excitation $h$ for several temperatures at fixed dc field $H$. Some of these measurements for the sample AD15 at $H=$5 Oe, are shown in \mbox{figure \ref{Xvshac}}. These curves clearly indicate the existence of a Campbell regime at low $h$ values, characterized by a susceptibility independent of $h$ (see dashed lines in the figure) and $\chi^{\prime \prime} \sim 0$. At higher amplitudes, the dissipation departs from zero and a crossover to an amplitude-dependent nonlinear regime takes place (see black arrows in this figure). 

\begin{figure}[htb]
\centering
\includegraphics[angle=0,width=90mm]{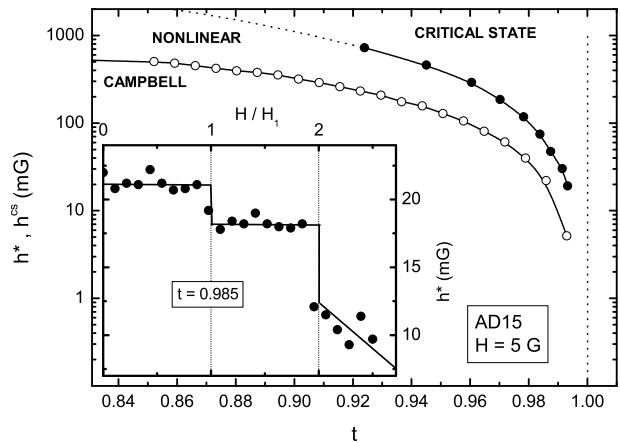}
\caption[]{{\small Main panel: dynamic $h-T$ phase diagram at $H=5$ G. The $h^*(T)$ curve represents the onset of the nonlinear response and $h^{cs}(T)$ indicates the beginning of the critical state regime. Inset: dynamic diagram in the $h-H$ plain at fixed temperature. The lines are guides to the eye.}}
\label{boundary}
\end{figure}

We use the criterion $\chi^{\prime \prime} = 0.01$ as a reliable estimation of the onset of the nonlinear regime. Performing this analysis for different temperatures and dc fields we determine the boundary $h^*(T,H)$ between the Campbell and the nonlinear regime. The obtained $h^*(t=T/T_c)$ for $H=5$ G, is shown in the main panel of Figure \ref{boundary} as open symbols. 

For drive fields $h < h^*$, the pinning potential can be approximated by a parabolic well and the ac-response arises from vortices performing small intra-valley oscillations inside their pinning site. For $h \agt h^*$, the average vortex displacement approaches the pinning range where the harmonic approximation breaks down and a nonlinear response is detected. Eventually, at higher amplitudes, vortices are able to hop out the potential well and a critical state develops.

Since the limits of the linear regime are given mainly by the strength of the pinning centers, it is expected that the extension of this regime decreases as temperature increases, in agreement with the observed behavior. Interestingly, in these patterned samples, we are able to substantially change the nature of the pinning potential, and thus its strength, by simply changing the external magnetic field. Indeed, for fields $H < H_1$ the density of pinning centers is larger than the density of vortices and each flux line is strongly pinned by an antidot. For $H > H_1$ extra vortices could still be attracted by the antidots if the maximum number of flux lines that an antidot can hold $n_s > 1$. In this case, since multiquanta vortices are weaker pinned than a single vortex, a reduction in the extension of the linear regime is expected at each matching field. For fields $H > n_s H_1$ the incoming vortices will sit in interstitial positions caged by the repulsive interaction of their strong pinned neighbors. This family of vortices, much weaker pinned and with higher mobility,\cite{rosseel} will dominate the ac-response and a more dramatic reduction of $h^*$ should occur.

This picture is consistent with the behavior of $h^*(H)$ shown in the inset of Figure \ref{boundary} at $T=7$ K ($t=0.985$). In this figure we observe that for $0 < H < H_1$, the $h^*$ data scatter around 21 mG (indicated by a straight line in the figure) whereas for $H_1 < H < H_2$ the average value drops to 18 mG. In both field ranges, the data points exhibit a similar dispersion showing small peaks near $H_1/2$ and $3H_1/2$, where the interstitial flux line lattice form a highly stable configuration. The small step in $h^*(H)$ at $H_1$ indicates the transition from single quantum pinning to double quanta pinning. At $H=H_2$, $h^*$ undergoes a stronger reduction down to 12 mG, followed by a different field evolution as $H$ increases. This more pronounced jump in $h^*$ suggests that for $H > H_2$, a further increase of the occupation number per hole is no longer energetically convenient and hence, incoming vortices will locate in interstitial sites (i.e. $n_s=2$). 

A first theoretical estimation of the maximum number $n_s$ of vortices trapped in a cylindrical cavity was done more than 30 years ago by Mkrtchyan and Schmidt\cite{schmidt}. According to this calculation, $n_s \approx b/4\xi(T)$. In our particular case, this gives $n_s \sim 1$. Although this value is smaller than the experimentally determined above, it is important to note that this model underestimates the real $n_s$ since it considers a single vortex line interacting with a cylindrical insulating defect with radius $b \ll \lambda$. Clearly, this hypothesis is not fulfilled in our system where $b \agt 2 \lambda(T)$. Additionally, it is also expected that $n_s$ increases as the applied dc field is increased.\cite{baert,doria,buzdin} An extension of the original work of Mkrtchyan and Schmidt to arbitrary large cavity radius has been recently done by  Nordborg and Vinokur\cite{nordborg} using the London approximation. Similar results were found by Buzdin\cite{buzdin} using the method of image vortices. This author showed that in a triangular vortex lattice, a two-quanta vortex becomes energetically favorable for temperatures such that $b^3<\xi(T)\lambda(T)^2$, a condition that, in our sample, is satisfied at $t<0.995$, in agreement with the experimental result.

\section{Critical State Regime}

As we pointed out above, for $h > h^*$, the vortex motion is no longer restricted to local oscillations around the pinning centers and vortex excursions may be larger than the distance between nearby pinning sites. The ac response in this case can be described within a critical state model. In the simplest scenario of a Bean model, the persistent current density $J$ is uniform and close to the critical value $J_c$, everywhere in the sample where the inter-valley motion takes place. In this regime, unlike the Campbell regime, the field penetration $\Lambda$ increases linearly with $h$ as,\cite{clem-sanchez}

\begin{equation}
\Lambda = \frac {c} {4\pi} \frac {h} {J(T)}
\label{eq:penetration}
\end{equation}

\begin{figure}[htb]
\centering
\includegraphics[angle=0,width=90mm]{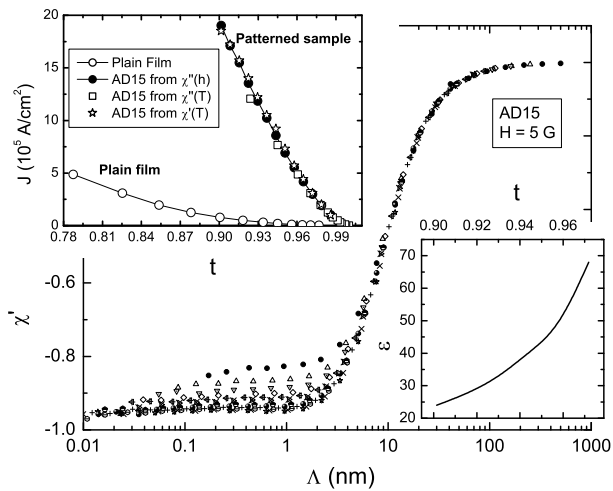}
\caption[]{{\small Screening $\chi^\prime$ as a function of the Bean penetration depth $\Lambda$ obtained by scaling the curves $\chi^\prime(h)$ showed in Figure \ref{Xvshac}. The upper inset shows the critical current as a function of temperature obtained from different methods for a plain film and a patterned film at $H=5$ G. The lower inset shows the efficiency $\varepsilon(T)=J(antidots)/J(plain film)$.}}
\label{Scaling-T}
\end{figure}

It is important to note that the ac response $\chi$ is solely determined by the penetration depth $\Lambda$ of the ac excitation. As a consequence, it is possible to find, in the critical state regime, a set of data at different $T$ and $h$, such that produce the same penetration $\Lambda$, and therefore the same $\chi$. For example, according to equation~(\ref{eq:penetration}), if all data in figure \ref{Xvshac} were in a well developed critical state regime, curves measured at constant $T$ should collapse on a single one when, for each curve, the horizontal axis $h$ is scaled by a suitable factor $J(T)^{-1}$. The resulting curve from this procedure is the Bean penetration depth $\Lambda(\chi^\prime) \approx h/J(T)$. This scaling is shown in the main panel of Figure \ref{Scaling-T} for temperatures ranging from 6 K to 7 K. We observe a good overlap at low screening values ($|\chi^\prime| < 0.7$) and a break down of the scaling when approaching to the Campbell regime at higher $\chi^\prime$ values, where the field penetration and therefore $\chi^\prime$, become $h$-independent. 

The set of parameters $J(T)$ chosen to obtain the superposition of the low screening data, determine the temperature dependence of the persistent current as is shown by star symbols in the upper inset of figure \ref{Scaling-T}. Alternatively, the persistent current $J(T)$ can be also obtained from $\chi^{\prime \prime}(h)$ curves measured at fixed $T$, by tracking the position $h^{max}(T)$ of the maximum dissipation. Indeed, exact calculations for the particular case of a disk show\cite{clem-sanchez} that $\chi^{\prime \prime}$ maximizes when $\Lambda=\delta$ and thus from equation~(\ref{eq:penetration}) we have that for each $T$, $J(T) \approx h^{max}/\delta$. This is shown as solid circles in the upper inset of \mbox{figure \ref{Scaling-T}}. A similar argument can be applied to obtain $J(T)$ from the temperature at which the dissipation maximizes in $\chi^{\prime \prime}(T)$ curves.\cite{vanderbeek91} The result of this procedure is shown by open squares in the same inset. 

For comparison, in the upper inset of \mbox{figure \ref{Scaling-T}} we also show $J(T)$ for a Pb plain film without antidots deposited simultaneously with the patterned film. The efficiency $\varepsilon(T) = J(antidots)/J(plain film)$ of the antidots as correlated pinning centers with respect to the random intrinsic pinning of the reference film, is shown in the lower inset of \mbox{figure \ref{Scaling-T}}. We observe that $\varepsilon$ monotonically decreases as $T$ decreases from $T_c$. This result is in agreement with the experimental fact that commensurability effects originated in the periodicity of the pinning array, progressively fade out as intrinsic pinning becomes more relevant at lower temperatures. The very high $\varepsilon$ values observed in this sample can be taken as a confidence test of the quality of the grown films.

Let us now analyze the crossover from the nonlinear intermediate regime to the critical state regime. According to the theory, in the simplest case of a cylinder in a parallel field, the critical state regime arises when the Campbell penetration depth $\lambda(T)$ becomes equal to the Bean penetration $\Lambda(T, h)$, i.e. when $h \sim J(T) \lambda(T)$.\cite{vanderbeek91} In a previous work\cite{silhanek03} we have determined for a similar sample, that $\lambda(t=0.90) \sim 250$ nm and using the $J(T)$ from \mbox{figure \ref{Scaling-T}}, we obtain a crossover field $h^{cs} \sim 40$ G, which turns out to be far above our experimental estimation of $h^{cs} \sim 1$ G according to \mbox{figure \ref{Scaling-T}}. Recently, Pasquini et al.\cite{pasquiniPRB} have pointed out that in the case of transverse geometry, since the actual range of field penetration does not coincide with the penetration depth, the previous criterion should be modified. Assuming that the range of field penetration in both regimes coincides when $|\chi^\prime| = 0.5$, the authors show that the crossover should occur at $h^{cs} \sim \sqrt{9\delta/2R} \lambda(T) J(T)$, where $R$ is the radius of a disk shaped sample. In our case, the geometric factor $\sqrt{9\delta/2R} \sim$ 0.007 shifts the crossover field down to $h^{cs} \sim$ 0.3 G. Although this crossover field is somewhat smaller than the experimental value, we should note that whereas this simplified model correctly accounts for the reduction of the transition field $h^{cs}$, it still involves a certain degree of arbitrariness since different criteria lead to different geometric factors. 

In fact, \mbox{figure \ref{Scaling-T}} shows that the scaling derived from the critical state becomes valid for $|\chi^\prime| < 0.7$ instead of $|\chi^\prime| = 0.5$. From the penetration depth at this screening value, $\Lambda_{cross} \sim 6$ nm, and the obtained critical current $J(T)$, we can estimate the onset of the critical state regime $h^{cs}=\Lambda_{cross} J(T)$. This boundary is represented by solid circles in the main panel of Figure \ref{boundary}. It is apparent in this figure that the transition between the linear and the critical state regime is not a sharp crossover but instead, a broad intermediate non linear regime lies in between. 

There are two different sources that give rise to this crossover regime. First, it appears as a natural consequence of the anharmonicity of the pinning potential. Second, due to the inhomogeneous current distribution, nonlinearities first appear at the border of the sample and then, as $h$ increases, the boundary separating inter- and intra-valley vortex motion, moves towards the center of the sample. This coexistence between linear and nonlinear regime gives rise to a nonlinear response. An additional effect that may further expand the intermediate regime is the presence of disorder in the distribution and energy of the pinning centers. However, as we pointed out above, in the studied samples owing to the regular array of pinning centers, the topological and energetic disorder is reduced down to minimum levels and should not play a significant role. 

\begin{figure}[htb]
\centering
\includegraphics[angle=0,width=90mm]{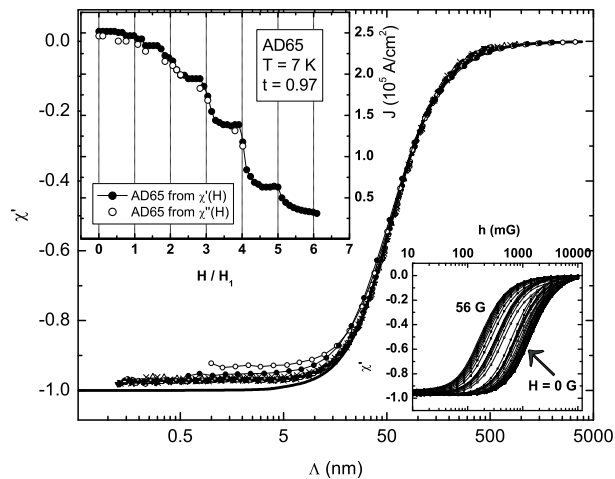}
\caption[]{{\small Screening $\chi^\prime$ as a function of the Bean penetration depth $\Lambda$ obtained by scaling the curves showed in the lower inset. The solid line depicts the theoretical expectations according to the Bean model. The lower inset shows the real part of the ac response $\chi^\prime$ as a function of the amplitude $h$ for several $H$ at $T=$ 7 K ($t=0.97$). The upper inset shows the critical current as a function of field at $t=0.97$, obtained following the procedures explained in the text.}}
\label{Scaling-H}
\end{figure}

A similar scaling analysis can be performed to obtain the field dependence of the critical current. To that end, we have measured the ac susceptibility response of the AD65 sample, as a function of $h$, for several applied dc fields $H$ at $T=7$ K ($t=0.97$). These curves are shown in the lower inset of \mbox{figure \ref{Scaling-H}} for fields ranging from 0 G to \mbox{56 G}. Strictly, equation~(\ref{eq:penetration}) is valid within the Bean critical state scenario, i.e. with a critical current $J_c$ independent of $H$.\cite{shatz} This shortcoming can be avoided either by applying an ac drive much smaller than the dc field, or provided that $J_c$ is insensitive to the small field oscillations.

In the main panel of \mbox{figure \ref{Scaling-H}} we show the results of a scaling procedure similar to that previously explained to obtain $\Lambda(\chi^\prime)$. For $|\chi^\prime|<0.6$, a fairly good overlap of all data points is clearly observed. This result indicates that in this field range, equation(\ref{eq:penetration}) is satisfied and therefore the Bean model applies. The solid line depicts the theoretical expectations according to the Bean model for a thin disk.\cite{clem-sanchez} It is worth to note that this is not a fitting curve but a theoretical prediction where no free parameters are involved.

The critical current $J(H)$ derived from this scaling is shown in the upper inset of \mbox{figure \ref{Scaling-H}} together with the values obtained from the position of the maximum dissipation $\chi^{\prime \prime}$. We observe that $J(H)$ exhibits barely defined crossover at $H_1$ and $H_2$ followed by more pronounced transitions at higher matching fields. It is important to note that, unlike the typical $\chi^\prime(H)$ measurements where several regimes are crossed while sweeping $H$, this scaling procedure allows one to estimate quantitatively the value of the critical current within a single dynamic regime.

\begin{figure}[htb]
\centering
\includegraphics[angle=0,width=90mm]{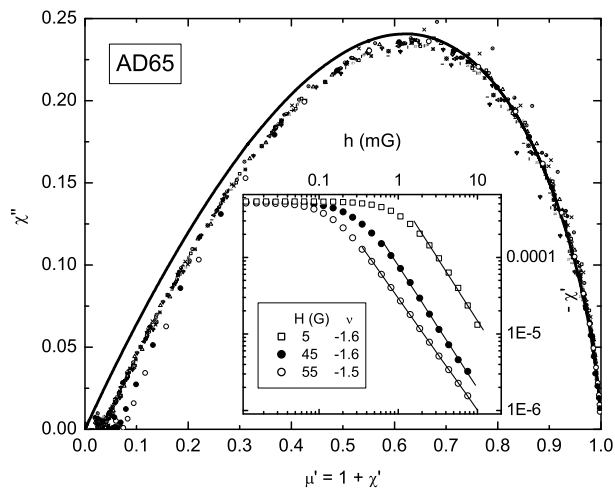}
\caption[]{{\small Experimental curves $\chi^{\prime \prime} (\mu^\prime)$ for the data points showed in \mbox{figure \ref{Scaling-H}}. The solid line corresponds to the theoretical calculation for a disk in the Bean critical state model. The inset shows the asymptotic behavior of $\chi^\prime$ for large $h$ together with the linear fits from where the exponent $\nu$ is determined (see text).}}
\label{ClemSanchez}
\end{figure}

Additional information on the dynamics in this regime can be obtained by plotting $\chi^{\prime \prime}$ vs. $\mu^\prime=1+\chi^\prime$ (Coles-Coles plot), as shown in \mbox{figure \ref{ClemSanchez}} at $t=0.97$ for several fields. Since in this kind of graph only dimensionless variables are considered, it results very suitable to analyze the data.\cite{review,shantsev,herzog} In this figure, the solid line represents the theoretical calculation of $\chi^{\prime \prime} (\mu^\prime)$ for a disk shaped sample in the Bean CS regime. For permeabilities  $|\mu^\prime|<0.1$ a very small dissipation indicates the presence of a linear regime. For higher $|\mu^\prime|$ values, the similarity between the experimental data and the theoretical curve strongly suggest the applicability of a Bean-like model at high $h$. On top of that we observe that the maximum of $\chi^{\prime \prime}$ is slightly smaller than the theoretically predicted, and it shifts towards $\mu^\prime \rightarrow 1$. These features indicate that flux creep is not relevant in this case, consistently with the observed weak $f$-dependence.

Further confirmation that data taken at large $h$ is well described by a Bean model can be obtained by analyzing the asymptotic behavior of the real part of the ac susceptibility $\chi^\prime$ at high $h$. Recently, Shantsev {\it et al.}\cite{shantsev} have shown that according to the CS model, $\chi^\prime \propto h^{-\nu}$ at large $h$, where $\nu$ depends on the $J(B)$ dependence. For example, in a Bean critical state model, $\nu=-3/2$, whereas for the exponential and the Kim model, $\nu=-3$. Accordingly, we have found, for the whole field range studied, that $\nu=-1.50 \pm 0.15$. Some of these curves are shown in the inset of \mbox{figure \ref{ClemSanchez}}. 

As we noted above, the average curvature of the pinning potential $\alpha$ obtained in the linear regime, together with the persistent current $J$ determined in the critical state regime, provide us with complementary information about the principal characteristics of the pinning landscape. Usually, $J$ is related with $\alpha$ through $\alpha u \approx J \Phi_0$, where $u$ is the pinning range. In this equation, the first term represent the restoring force due to the parabolic pinning potential and the second term is the Lorentz force exerted by the current $J$. We can use this expression in order to estimate the pinning range $u$ from the two independently obtained parameters $\alpha$ and $J$. In particular, taking the values $\alpha(t=0.90) \approx 5 \times 10^3$ G$^2$ from Ref.[\onlinecite{silhanek03}], and the current density $J(t=0.90) \approx 5 \times 10^5$ A/cm$^2$ for that sample, we obtain a pinning range $u \sim 330$ nm which is consistent with the expected range $u \sim \lambda \approx 400$ nm.

Finally we would like to point out that alternative models other than a critical state have been proposed for this kind of systems. In particular, Cooley and Grishin\cite{cooley} have shown that in a one-dimensional system with a periodic pinning array, a terraced flux profile should be established. Similarly, later on Reichhardt et al.,\cite{reichhardt-97} by means of molecular dynamic simulations, showed that in a 2D system, magnetic flux penetrate in the sample forming complex patterns like islands and striped domains. Although clearly, further theoretical works for higher dimensions and considering realistic sample geometries are needed, it is important to stress that the observed ac response can be satisfactorily accounted for by a simple Bean model without the necessity of invoking a multi-domain model. 

\section{Conclusions}

In summary, we have demonstrated that the Bean critical state model accurately describes the scaling observed in the ac susceptibility response and provides a reliable way to determine the critical current of the system as a function of temperature and field. The information gathered from different dynamic regimes allowed us to determine the averaged pinning range of the periodic landscape and estimate the boundary between the Campbell and the critical state regimes.

\section*{Acknowledgements}
We would like to thank R. Jonckheere for fabrication of the resist pattern. This work was supported by the Belgian Interuniversity Attraction Poles (IUAP), and Research Fund K.U.Leuven GOA/2004/02, the Fund for Scientific Research Flanders (FWO) and ESF ``VORTEX'' program. AVS and MJVB are Posdoctoral Research Fellows of the FWO.

\end{document}